\documentclass[letterpaper, 10 pt, conference]{ieeeconf}
\usepackage{amsmath,amssymb,paralist,subfigure,graphicx,color,stmaryrd,mathrsfs,url,cite}
\usepackage{siunitx}
\usepackage[ruled,vlined,linesnumbered]{algorithm2e}\usepackage[table]{xcolor}

\usepackage{blkarray}
\usepackage{bbm}
\usepackage{bm}
\usepackage[makeroom]{cancel}
\usepackage{cases}
\usepackage{dsfont}
\usepackage{epsfig}
\usepackage[T1]{fontenc}
\usepackage{graphicx}
\usepackage{flushend}
\usepackage{textcomp}
\usepackage{xcolor}
\usepackage[UKenglish,english]{babel}
\usepackage{amsmath,amssymb,paralist,subfigure,graphicx,stmaryrd,mathrsfs,url,algorithm2e,soul}
\usepackage{cuted,mathtools,lipsum,cite}
\usepackage{float}
\usepackage{flushend}
\usepackage{graphicx}
\usepackage{hyperref} 
\usepackage{mathtools}
\usepackage{mathrsfs}
\usepackage{setspace}
\usepackage{soul} 
\usepackage{subfigure}
\usepackage{stfloats}
\usepackage{tikz}
\usepackage{verbatim}
\usepackage{xspace}

\newtheorem{lem}{Lemma}
\newtheorem{ass}{Assumption}

\newtheorem{rem}{Remark}

\def\mb{\mathbf}

\def\mc{\mathcal}

\IEEEoverridecommandlockouts                
\begin{document}
\title{\LARGE \bf Distributed Anomaly Detection and Estimation over Sensor Networks: Observational-Equivalence and $Q$-Redundant Observer Design}
\author{Mohammadreza Doostmohammadian,  Themistoklis Charalambous,~\IEEEmembership{Senior Member,~IEEE}
	
\thanks{Authors are with the School of Electrical Engineering at Aalto University, Espoo, Finland, Email: \texttt{firstname.surname@aalto.fi}. M. Doostmohammadian is also with the Faculty of Mechanical Engineering at Semnan University, Semnan, Iran, Email: \texttt{doost@semnan.ac.ir}. T. Charalambous is also with the Electrical Engineering Department at the University of Cyprus, \texttt{surname.name@ucy.ac.cy}. }}
\maketitle

\begin{abstract}
	In this paper, we study stateless and stateful physics-based anomaly detection scenarios via \textit{distributed estimation} over sensor networks. 
	In the stateful case, the detector keeps track of the sensor residuals (i.e., the difference of estimated and true outputs) and reports an alarm if certain statistics of the recorded residuals deviate over a predefined threshold, e.g.,  $\chi^2$ (Chi-square) detector. Instead, only instantaneous deviation of the residuals raises the alarm in the stateless case without considering the history of the sensor outputs and estimation data. Given (approximate) false-alarm rate for both cases, we propose a probabilistic threshold design based on the noise statistics. We show by simulation that increasing the window length in the stateful case may not necessarily reduce the false-alarm rate. On the other hand, it adds unwanted delay to raise the alarm. The distributed aspect of the proposed detection algorithm enables local isolation of the faulty sensors with possible recovery solutions by adding redundant observationally-equivalent sensors. We, then, offer a mechanism to design $Q$-redundant distributed observers, robust to failure (or removal) of up to $Q$ sensors over the network.
	
\end{abstract}

\section{Introduction} \label{sec_intro}
The recent advancements in wireless communication, high-performance networking, sensing, and processing devices, along with cloud-computing and Internet-of-Thing (IoT)  \cite{csl_iot,chen2018internet} have motivated localized (or \textit{distributed}) estimation over sensor networks.  Such localized setup, further, mandates \textit{distributed detection} mechanisms to find possible anomalies (faults, failures, or malicious attacks) \textit{locally}, with potential large-scale applications from social networks \cite{icas21_social} to Cyber-Physical Energy Systems (CPES) 
\cite{camsap11,ilic2010modeling}. 
Such distributed methods are privileged with \textit{no single node of failure}  and outperform existing centralized detection  \cite{umsonst2019tuning,abbaszadeh2019system,davoodi2016event,navi2018sensor}.


The existing distributed methods either require (i) \textit{local} system observability in the neighborhood of every sensor via high-traffic communication network with large quantity of data-transfer \cite{chen2018internet}, or (ii) fast data-sharing and processing units to perform many iterations of consensus and communication between every two samples of  system dynamics (\textit{double time-scale}) \cite{he2020secure,battilotti2021stability}. In this work, similar to \cite{tcns20,morse2019cdc,zamani2018distributed}, we consider \textit{single time-scale} distributed estimation (i.e., iterations at the same time scale of system dynamics) with no assumption on local system observability at any sensor \cite{jstsp,camsap11} to relax the networking, communication, and computational needs. A challenge is to build distributed estimation networks robust to faults, anomalies, and even cyber-attacks. Such detection and mitigation techniques are wide-spread in \textit{centralized} setup \cite{giraldo2018survey}. Some works propose preventive mechanisms as in privacy-preserving  \cite{rikos2020privacy}, and resilient estimation \cite{sundaram_2021resilient,kim2018detection}, while many other use detection mechanisms, e.g., observer-based (or predictive) fault detection and isolation (FDI)  \cite{davoodi2016event,navi2018sensor}.
Machine learning and binary classification algorithms (e.g., support-vector-machine \cite{dsvm}) are also used to generate decision boundaries separating \textit{normal space} of system states from \textit{abnormal space} (operating with fault or anomaly) \cite{abbaszadeh2019system}. 
Graph-theoretic methods via \textit{structured systems theory} are also adopted for \textit{generic} design of detection/mitigation protocols irrespective of numerical parameter values and only based on system \textit{structure} \cite{asilomar14,GRACY2021109925}. 
What missing in the literature is a \textit{distributed} algorithm to \textit{localize} the detection, in both stateful and stateless setups, along with network design for \textit{redundant distributed observability} (distributed fault-tolerant models). 

\textit{Main contributions:}
We propose a local mechanism to detect possible bias/anomaly in system output while performing single time-scale distributed estimation. We design probabilistic detection thresholds in both stateless and stateful scenarios, define the false-alarm (false-positive) rate (FAR) and false-negative rate for each case, and compare the performance of both solutions in terms of FAR (and delay in raising the alarm). Further, in contrast to the full-rank model in \cite{icas21_social,tcns20}, this work considers distributed detection on (possibly) rank-deficient systems, known to require more information-sharing over the distributed estimation networks \cite{jstsp}. We specifically use constrained LMI gain design to \textit{isolate} (faulty) sensor residuals. Another contribution is to design $Q$-redundant distributed observers, robust to failure or removal of any $Q$ (failed or faulty) sensors. Using graph-theory (structural methods), we first define the set of \textit{observationally-equivalent} sensors/state-outputs using \textit{strongly-connected-components} (SCCs) and \textit{contractions} in the \textit{system digraph}. Then,  using $k$-vertex-connected (or $k$-connected) graphs \cite{guichard2017introduction}, we propose sensor-network structures resilient to  $Q$ sensor removals such that the remaining sensor-network successfully tracks the underlying system (which is not necessarily observable in the neighborhood of any sensor). The proposed algorithms are of polynomial-order complexity.


\textit{Paper organization:} the problem is set up  in Section~\ref{sec_fram}. The distributed detection mechanism  and $Q$-redundant observer are proposed in Section~\ref{sec_main} and~\ref{sec_Q}. The simulations and conclusion are given in Section~\ref{sec_sim} and \ref{sec_con}.  

\section{The Framework} \label{sec_fram}
\subsection{System model}
In this paper, we consider LTI systems
in the form,
\begin{align}\label{eq_sys1}
\mb{x}(k+1) &= A\mb{x}(k) + \pmb{\nu}(k),
\end{align}
with $k$ as the time index, $A$ as the system matrix (\textit{possibly rank-deficient}), ${\mb{x} \in \mathbb{R}^n}$ as the column-vector of states,  i.e., ${\mb{x}=[x_1;\cdots;x_n]}$,  
and ${\pmb{\nu} \sim \mc{N}(0,\Sigma_\nu)}$ as the Gaussian noise. The output vector at time $k$ is in the form,
\begin{align}
\mb{y}(k) &= C\mb{x}(k) + \pmb{\zeta}(k)+\mb{f}(k), \label{eq_C}
\end{align}
with $\mb{y}=[y_1,\cdots,y_N] \in \mathbb{R}^N$, $\pmb{\zeta} \sim \mc{N}(0,\Sigma_\zeta)$ as the output noise (with diagonal $\Sigma_\zeta$), and $\mb{f} \in \mathbb{R}^N$ as the \textit{additive bias} to the outputs (due to faults or anomalies) defined as,
\begin{align}
     \left\{
  \begin{array}{@{}lr}
     f_i(k) = 0 & \mbox{no-fault at output } i \\
     f_i(k) \neq 0 & \mbox{faulty output } i
  \end{array}\right.
\end{align}
Without loss of generality, in this paper, we assign every sensor $i$ with one state-output $y_i$. 
Given the structural representation of $A,C$ matrices (i.e., the \textit{system digraph} $\mc{G}_A$), sufficient conditions for structural (or \textit{generic}) $(A,C)$-observability are given in \cite{jstsp,acc13_kar}. Such structural methods imply  observability for \textit{almost all} numerical values of non-zero system parameters. This is referred to as the \textit{linear-structure-invariant} (LSI) model, where the structure is fixed and the non-zero entries in $A$ and $C$ can take almost any value (with non-admissible parameter values lying on an algebraic subspace with \textit{zero Lebesgue measure}) \cite{jstsp}. 

\subsection{Distributed consensus-based estimation}
We consider a distributed estimation framework, where the (group of) sensors estimate the system $A$ locally via information-sharing over a sensor network to gain \textit{distributed observability} (as defined later in Lemma~\ref{lem_dist_obs}).
The following distributed estimator in \textit{single time-scale}, proposed in \cite{jstsp}, is considered at sensor $i$.

\small \begin{align}\label{eq_p} 
\widehat{\mb{x}}_i(k|k-1) =& \sum_{j\in\mathcal{N}_\beta(i)} W_{ij}A\widehat{\mb{x}}_j(k-1|k-1),
\\ \nonumber
\widehat{\mb{x}}_i(k|k) =& \widehat{\mb{x}}_i(k|k-1) \\
&+ K_i \sum_{j\in \mc{N}_\alpha(i)}C_j \left({y}_j(k)-C_j^\top \widehat{\mb{x}}_i(k|k-1)\right), \label{eq_m}
\end{align} \normalsize
where  $\widehat{\mb{x}}_i(k|k-1)$ and $\widehat{\mb{x}}_i(k|k)$ denote the priori and posteriori estimates via the received data up to time $k-1$ and $k$, respectively. $K_i$ is the local gain (to be designed), and $\mc{N}_\alpha(i)$ and $\mc{N}_\beta(i)$ denote the in-neighborhood  respectively over networks $\mc{G}_\alpha$ and $\mc{G}_\beta$ (defined in Lemma~\ref{lem_dist_obs}),  with 0-1 matrix $U=[U_{ij}]$  and \textit{row-stochastic} matrix $W=[W_{ij}]$. Recall that $\mc{N}_\alpha(i)$ includes the so-called $\alpha$-sensors with outputs of  rank-deficient part of the system \cite{jstsp,asilomar14}, where for $j\in \mc{N}_\alpha(i)$ the entry $U_{ij}=1$. The following assumption distinguishes this work from many literature
on single time-scale distributed estimation and observer-based detection.
\begin{ass} \label{ass_obs}
The pair $(A,C)$ is observable and the pair $(A,C_{j\in \mc{N}_{\alpha}(i)})$ (with $C_{j\in \mc{N}_{\alpha}(i)}$ representing outputs in  $\alpha$-neighborhood of $i$) is \textit{not necessarily observable} at any sensor $i$. In other words, we assume \textit{global observability} at the group of \textit{all sensors} and \textit{no local observability} in the \textit{neighborhood of any sensor}.
\end{ass}
\begin{rem}
The proposed single time-scale protocol, similar to \cite{tcns20,morse2019cdc,zamani2018distributed}, is more suitable for large-scale, as compared to  double time-scale methods \cite{he2020secure,battilotti2021stability}, in terms of needed communication traffic and computation loads on sensors. This is because, in the latter, sensors perform a large number of consensus/data-sharing iterations between steps $k$ and $k+1$ of system dynamics, in contrast to only $1$ iteration in single time-scale methods. See details in \cite{camsap11,jstsp,tcns20}.
\end{rem}

 Define the estimation error as $\mb{e}_i(k)= \widehat{\mb{x}}_i(k|k)-\mb{x}(k)$ and $\mb{e}=[\mb{e}_1,\cdots,\mb{e}_N]$. The error dynamics under \eqref{eq_p}-\eqref{eq_m} is,
  \normalsize
\footnotesize
\begin{align}\label{eq_err}
\mb{e}(k+1) &= (W \otimes A)\mb{e}(k) -K D_C(W \otimes A)\mb{e}(k)+\pmb{\eta}(k+1), \\ \label{eq_eta}
\pmb{\eta}(k+1)  &= (I-K D_C)\mb{1}_N \otimes \nu(k) -K \overline{D}_C (\pmb{\zeta}(k+1) +\mb{f}(k+1))
\end{align}\normalsize
with $\pmb{\eta}(k+1)$ containing the noise and fault terms, $nN\times nN$ matrices $D_C=\mbox{diag}[U_{ij}C_j C_j^\top]$ and $\overline{D}_C \triangleq (U \otimes \mb{1}_n) \circ (\mb{1}_N \otimes C^\top) $ ("$\circ$" and "$\otimes$" respectively as the entry-wise  and Kronecker product). Let $\widehat{A} := W \otimes A-K D_C(W \otimes A)$. Eq.~\eqref{eq_err} represents a cumulative LTI dynamics with system matrix $\widehat{A}$, where $\rho(\widehat{A})$ determines the error stability (with $\rho(\cdot)$ as the spectral radius). Following the Kalman stability theorem, one can design block-diagonal gain matrix $K=K \circ \mc{K}=\mbox{diag}[K_i]$ (with $ \mc{K}=I_N \otimes \mb{1}_{n\times n}$, $I_N$ and $\mb{1}_{n\times n}$ respectively as identity and ones matrix of size $N$, $n$) to stabilize~\eqref{eq_err} \textit{if the pair $(W \otimes A, D_C)$ is observable}. This is referred to as the distributed observability, with the sufficient conditions discussed in the next lemma. 

\begin{lem}[\cite{jstsp}]
\label{lem_dist_obs}
The pair $(W \otimes A, D_C)$ is observable if (i) $\mc{G}_\beta$ is strongly-connected, and (ii) every $\alpha$-sensor is a \textit{hub} of $\mc{G}_\alpha$, where every sensor $i$ with a system-output recovering  \textit{structural rank-deficiency}\footnote{Later in Section~\ref{sec_Q}, using the system graph representation, an $\alpha$-sensor is defined as the sensor with output of an state node in a contraction component. We refer interested readers to \cite{icassp2016} for more details.} is regarded as an $\alpha$-sensor. 
\end{lem}

Then, the block-diagonal gain matrix $K$ can be designed, e.g., via the  LMI in \cite{jstsp,usman_cdc:11}, such that the error dynamics is Schur stable, i.e., $\rho(\widehat{A})<1$ for general (possibly unstable) $A$. Define the steady state error variance as
$\Sigma_e =  \lim_{k\rightarrow \infty} \mathbb{E}(\mb{e}_i(k)^\top\mb{e}_i(k))$. It is shown in \cite{jstsp,usman_cdc:11} that $\lim_{k\rightarrow \infty} \mathbb{E}(\mb{e}_i(k))=\mb{0}$, and,
\begin{align} \label{eq_sigma_e}
    \|\Sigma_e\|_2 = \frac{a_1N\|\Sigma_\nu\|_2+a_2 \|\overline{\Sigma}_\zeta\|_2}{1-b^2},
\end{align}
with $b :=\|(W\otimes A) - K D_C (W\otimes A) \|_2<1$, $a_1 := \|I_{Nn}- K D_C\|_2^2$, $a_2 := \|K\|_2^2$, $\overline{\Sigma}_\zeta := \mbox{diag}[\sum_{j\in \mc{N}_\alpha(i)} C_j^\top \Sigma_\zeta^j C_j]$.
where $\Sigma_\zeta^j$ denotes the $j$th diagonal entry of $\Sigma_\zeta$.

\section{Main Results: Local Detection of Anomalies} \label{sec_main}
\subsection{Stateless Detector}
\begin{figure}
	\centering
	\includegraphics[width=2.75in]{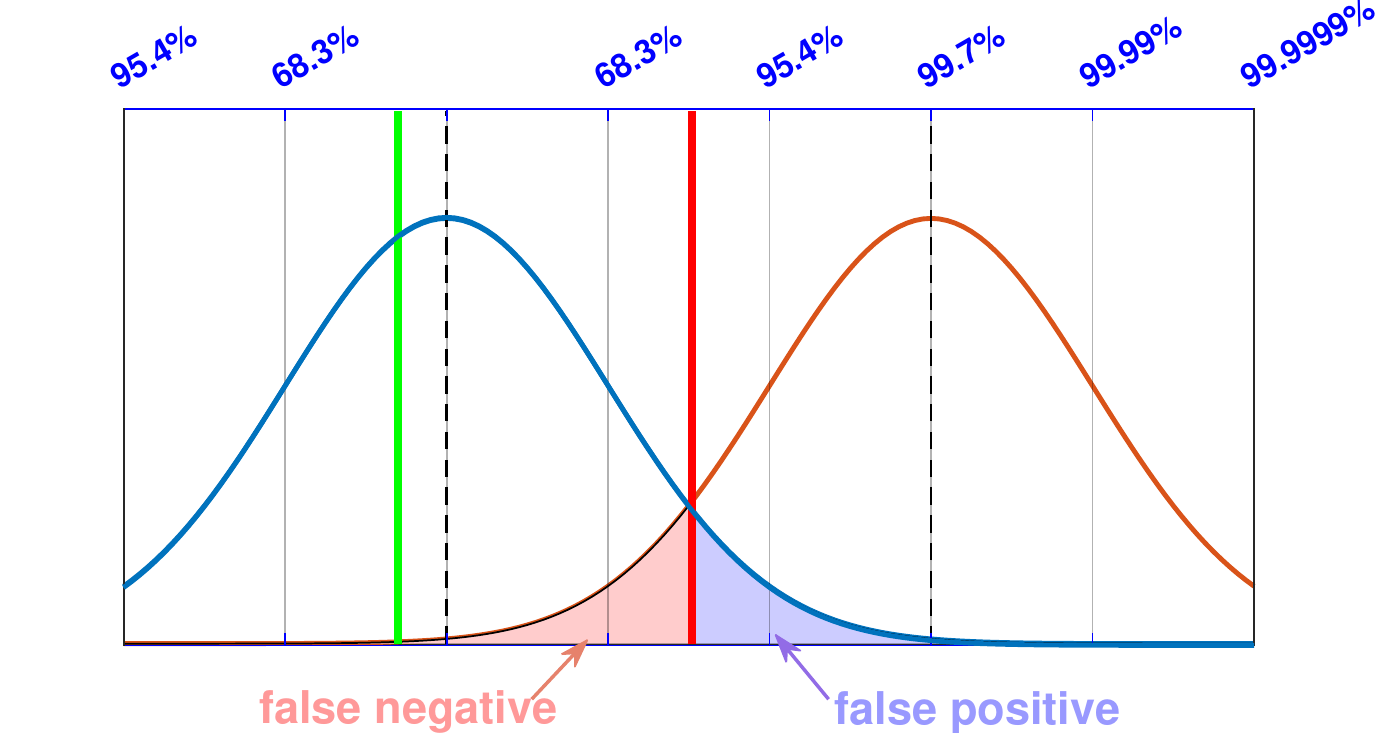}
	\caption{This figure shows example distributions for unbiased residual with $f_i(k)=0$ (blue curve) versus the biased residual with $f_i(k)\neq 0$ (red curve). The confidence intervals (of the unbiased PDF) are shown with the associated probabilities. Two vertical green and red lines represent two example residual values $r_k^i$, where the green one (most likely) belongs to the unbiased curve as it is close to the expected value (zero) of the unbiased PDF and, thus, most likely is due to system/measurement noise; the red sensor residual, however, is far from the expected value (zero) and, based on the shown confidence-intervals, represents anomalous output with FAR $4.6\% < p < 31.7\%$ (more accurately $p=1-\mbox{erf}(\frac{\kappa = 1.5}{\sqrt{2}})=13.3\%$ with $\theta_p = 1.5 \Sigma_r^i$). Considering the absolute residual $|r_k^i|$ and folding-over the red PDF to the right-hand-side (RHS) of $0$ axis, the blue shaded area equals to $\frac{p}{2}$, representing half of the probability of false-alarm, where the other half on the LHS of $0$ is not shown for simplicity. Similarly, the red shaded area, equal to $\frac{p}{2}$, \textit{approximately} represents the false-negative rate.} \vspace{-0.6cm}
	\label{fig_normal} 
\end{figure}

Given the error dynamics \eqref{eq_err}-\eqref{eq_eta}, define the residual of sensor $i$ (including possible anomaly/bias $f_i$) at time $k$ as, 
\begin{align} \nonumber
    r_i(k) &= y_i(k)- C_i\widehat{x}_i(k|k)= C_i\mb{e}_i(k)+\zeta_i(k)+f_i(k) \\  \label{eq_rbar2}
    &= C_i \widehat{A}_i \mb{e}_i(k-1)+ C_i \pmb{\eta}_i(k) +\zeta_i(k)+f_i(k) 
\end{align}
with 
\footnotesize
\begin{align} \nonumber
    \zeta_i&(k)+f_i(k)+C_i \pmb{\eta}_i(k)  = \zeta_i(k)+f_i(k)+C_i \pmb{\nu}(k-1)\\ \label{eq_eta_f}
    &-C_i K_i \sum_{j\in \mc{N}_\alpha(i)}\left( C_j^\top \zeta_j(k)+C_j^\top f_i(k) +  C_j C_j^\top \pmb{\nu}(k-1) \right)
\end{align} \normalsize
In steady-state (large enough $k$),  the noise and (non-zero) fault terms in \eqref{eq_eta_f} mainly value the residual of sensor $i$. From \eqref{eq_eta_f}, other than bias $f_i$, any fault in the $\alpha$-neighborhood of sensor $i$, say $f_j,j\in \mc{N}_\alpha(i)$, also appears in $r_i(k)$. This makes the fault isolation  more challenging as compared to the full-rank system models \cite{tcns20,icas21_social}  in which  $\mc{N}_\alpha(i)=\{i\}$. To overcome this and  isolate the faults on $j\in\mc{N}_\alpha(i),j\neq i$, from \eqref{eq_eta_f}, we constrain the gain matrix $K_i$ such that,
\begin{align} \label{eq_const_K}
    |C_i^\top K_i C_j | \leq \epsilon |1 - C_j^\top K_j C_j |, \forall j \in \mc{N}_\alpha(i), j \neq i
\end{align}
with $\epsilon<1$ as a design constant, implying that the fault-related term $C_i^\top K_i C_j f_j$  in the residual $r_i(k)$ due to fault $f_j$ is down-scaled by $\epsilon$ as compared to the fault-related term in  $r_j(k)$ (of $\alpha$-sensor $j$ itself). The  constraint~\eqref{eq_const_K} helps to isolate possible faults at $\alpha$-sensors in rank-deficient systems via Algorithm~\ref{alg_lmi} as the modified version of the LMI gain design in \cite{tcns20}. This algorithm can be run either once centralized offline with $K_i$s given to the nodes after termination or iteratively over $k$.
\begin{algorithm} [t] \label{alg_lmi}
	\textbf{Input:} matrices $A$, $W$, $U$, $C$, scale factor $\epsilon$ \\
    Calculate $D_C=\mbox{diag}[U_{ij}C_j C_j^\top]$ \; 
    \If{ $(W \otimes A,D_C)$ is not observable}
    {Terminate\;}
    \Else{Iteratively solve for $\widehat{A} = W\otimes A - KD_C(W\otimes A)$ 
\begin{align}
\begin{aligned}
\displaystyle
\min
~~ &  \mathbf{trace}(XY) \\
\text{s.t.}  ~~& X,Y\succ 0,  & K \leftarrow  K \circ \mc{K} 
\\ \nonumber ~ & \left( \begin{array}{cc} X&\widehat{A}^\top\\ \widehat{A}&Y\\ \end{array} \right) \succ 0,~& \left( \begin{array}{cc} X&I\\ I&Y\\ \end{array} \right) \succ 0,
\end{aligned}\\ \nonumber
\frac{|C_i K_i C_j^\top|}{|1-C_j K_j C_j^\top |}<\epsilon,~\forall j \in \mc{N}_\alpha(i),~j \neq i
\end{align}
}
	\textbf{Output} Block-diagonal gain matrix $K =  K \circ \mc{K}$\;	
	\caption{Constrained iterative LMI gain design} 
\end{algorithm}
From \eqref{eq_sigma_e} and results in \cite{khan2014collaborative,tcns20}, one can define  the \textit{confidence intervals} for $r_i(k)$. First, note that from \eqref{eq_rbar2} for the case of no anomaly/fault $f_i(k)=0,\forall i$ and,
\begin{align} \label{eq_sigma_r}
    \Sigma_r^i \leq  \frac{\|C_i\|_2 \|\Sigma_e\|_2}{N} + \Sigma_\zeta^i
\end{align}
where, assuming one state measurement by every sensor $i$, $\|C_i\|_2 = |C_i|$.  Then, the probability that the residual lies in the confidence interval  $|r_i(k)| \leq \kappa \Sigma_r^i$ is ${\mbox{erf}(\frac{\kappa}{\sqrt{2}})}$ with $\mbox{erf}(\cdot)$ as the \textit{Gauss error function}. One can similarly claim that the probability of $|r_i(k)| \geq \kappa \Sigma_r^i$ is $1-\mbox{erf}(\frac{\kappa}{\sqrt{2}})$ \textit{for zero additive bias} $f_i(k)=0$. Define the \textit{probabilistic threshold} $\theta_p = \kappa \Sigma_r^i$. Then, the detection logic in this paper is as follows: for given FAR (or false-positive probability) $p$ define the threshold $\theta_p$ with  $\kappa=\sqrt{2}\mbox{erf}^{-1}(1-p)$. If $|r_i(k)| \geq \theta_p$ raise the anomaly detection alarm  (associated with FAR $p$). Recall that we use \textit{absolute residual} $|r_i(k)|$, and thus, \textit{folded} Gaussian distribution, i.e., to fold-over the probability mass to the RHS of $0$ axis by taking the absolute value. Similarly, the probability of \textit{false negative} can be defined as,

\footnotesize
\begin{align} \label{eq_false_n}
    \overline{p} &= \frac{1-\mbox{erf}(\kappa/\sqrt{2})}{2}-\frac{1-\mbox{erf}(3\kappa/\sqrt{2})}{2} 
    = \frac{\mbox{erf}(3\kappa/\sqrt{2})-\mbox{erf}(\kappa/\sqrt{2})}{2}
\end{align} \normalsize
For $\kappa\geq2$, we have $1-\mbox{erf}(3\kappa/\sqrt{2}) \approx 2\mathrm{e}{-10}$ and one can approximate $\overline{p}$  by $\frac{p}{2}$. This is better explained in Fig.~\ref{fig_normal}.
This detection logic  uses the instant value of the $r_i(k)$ at every time $k$ with no use of residual history, known as the \textit{stateless} mechanism. Next, we consider  residual history for anomaly detection, which is known as the \textit{stateful} mechanism.

\subsection{Stateful Detector}
Define \textit{distance measure} $\iota_i(k)$ as,
\begin{equation} \label{eq_z}
    \iota_i^T(k) = \sum_{m=k-T+1}^k \frac{r_i(m)^2}{\Sigma_r^i}.
\end{equation}
over a \textit{sliding time-window} of length $T$. It is known that the summation of squared random variables from normal distribution (i.e., $\frac{r_i(k)^2}{\Sigma_r}$) follows the so-called \textit{Chi-squared} distribution with $T$ degrees of freedom (denoted by $\chi^2_T$) \cite{giraldo2018survey,icas21_social,Bausch2013OnTE}, where $\mathbb{E}(\iota_i^T)=T$. Following the same line of reasoning as in the stateless case, the probabilistic threshold on  the variable $\iota_i^T$ (for given FAR $p$) can be defined as,
\begin{align} \label{eq_theta_T}
    \theta_p^T = 2\Gamma^{-1}(1-p,\frac{T}{2})
\end{align}
where $\Gamma^{-1}(\cdot,\cdot)$ denotes the \textit{inverse regularized lower incomplete gamma function}.
\textit{Weighted} distance measure  \cite{umsonst2019tuning} further can be considered to put more weights on the recent (normalized) residuals and less on the far past residuals as,
\begin{equation} \label{eq_z2}
    \overline{\iota}_i(k) = \sum_{m=k-T+1}^k \mu^{k-m} \frac{r_i(m)^2}{\Sigma_r^i}.
\end{equation}
with $0<\mu \leq 1$ as the weight factor. $\overline{\iota}_i$ is referred to as the weighted sum of Chi-squared distributions \cite{Bausch2013OnTE}. Although, $\chi^2$-distribution is typically defined for positive integer $T$, one can find similar results for the weighted $\chi^2$ given by \eqref{eq_z2} \cite{Bausch2013OnTE}, with the expected value of $\overline{\iota}_i$ and the probabilistic thresholds (for a pre-specified FAR $p$) similar to \eqref{eq_theta_T}, 
\begin{align} \label{eq_theta_T2}
    \mathbb{E}(\overline{\iota}_i) = \frac{1-\mu^{T}}{1-\mu},~\theta_p^\mu = 2\Gamma^{-1}\left(1-p,\frac{1-\mu^{T}}{2-2\mu}\right)
\end{align}
For the stateful detectors \eqref{eq_theta_T} and \eqref{eq_theta_T2},  the FAR is a function of $T$ and $\mu$ using the CDF of the $\chi^2$-distribution,
\begin{align}
    p = 1-\frac{\gamma (\frac{\iota_i^T}{2},\frac{T}{2})}{\Gamma(\frac{T}{2})},~p = 1-\frac{\gamma (\frac{\overline{\iota}_i}{2},\frac{1-\mu^{T}}{2-2\mu})}{\Gamma(\frac{1-\mu^{T}}{2-2\mu})} \label{eq_p_Tmu}
\end{align}
with $\gamma(\cdot,\cdot)$ as the \textit{lower incomplete gamma function}. Our fault detection logics are summarized in Algorithm~\ref{alg_logic} with either stateless (first \texttt{if}) or stateful detection (second \texttt{if}).  

\begin{rem} \label{rem_mu}
   For the stateful case, longer time-window $T$ results in less FAR; however, it also increases the false alarm \textit{delay} \cite{giraldo2018survey}. A similar statement holds for the weight factor $\mu$; greater $\mu$  results in lower FAR and more-delayed alarm. In general, there is a trade-off between detection accuracy  (lower FAR via increasing $T$ and $\mu$) and detection delay in raising the alarm; see examples in Section~\ref{sec_sim}. 
\end{rem} 
\begin{algorithm} [t] \label{alg_logic}
	\textbf{Input:} $A$, $W_{ij}$, $U_{ij}$, $C_i$, $y_i$ at sensor $i$, FAR $p$, $K_i$ via Algorithm~\ref{alg_lmi},  time-window $T$, weight-factor $\mu$ \\
    Find $r_i(k)$ via \eqref{eq_rbar2}, $\iota_i^T(k)$ via \eqref{eq_z}, or $\overline{\iota}_i(k)$ via \eqref{eq_z2} at step $k$\;
    Define $\theta_p = \kappa \Sigma_r^i$, $\theta_p^T$ via \eqref{eq_theta_T}, or  $\theta_p^\mu$ via \eqref{eq_theta_T2}\;
    \Begin(Stateless detection:){
    \If{ $r_i(k)\geq \theta_p$}
    {Declare: alarm at sensor $i$\;}}
    \Begin(Stateful detection:){
    \If{ $\iota_i^T(k) \geq \theta^T_\kappa$ or $\overline{\iota}_i(k) \geq \theta^\mu_\kappa$}
    {Declare: alarm at sensor $i$\;}
}
	\textbf{Output} Binary decision (Alarm or no-Alarm) associated with pre-specified FAR $p$\;
	\caption{Localized Detection of Output Bias}
\end{algorithm}
\begin{rem}
   The proposed detection logic in Algorithm~\ref{alg_logic} is \textit{localized} at every sensor $i$ with no need of centralized decision making. This is more feasible in large-scale applications as compared to the centralized detection methods \cite{abbaszadeh2019system,davoodi2016event,navi2018sensor}.
\end{rem}

\section{Q-redundant distributed observer design  based on structural observational equivalence} \label{sec_Q}
In this section, we provide $Q$-redundant version of the distributed estimator \eqref{eq_p}-\eqref{eq_m} such that it tolerates removal (or isolation) of any $Q$ (faulty) sensors while holding distributed observability over the remaining sensor-network. Thus, the other sensors can locally estimate the system and detect possible output bias/anomaly. This problem is twofold: (i) it provides sufficient outputs such that after removal of up to $Q$ rows of matrix $C$, the pair $(A,C)$ remains structurally observable (known as $Q$-redundant observability \cite{kim2018detection}), and (ii) designs $Q$-redundant communication network such that the conditions in Lemma~\ref{lem_dist_obs} hold after removing $Q$ sensors and cutting their linking over the network. We address these via the notion of observational-equivalence \cite{icassp2016,icas21_contraction}. 

\subsection{$Q$-redundant observability} \label{sec_Q1}
Given an observable pair $(A,C)$, two outputs $y_i=C_i\mb{x}$ and $y_j=C_j\mb{x}$ are \textit{observationally-equivalent} if losing either of the two does not affect system observability, while removing both makes the system \textit{unobservable}.  Let $\overline{C}_i$ denote the output matrix after removing the row $C_i$ from $C$ (i.e., removing sensor $i$). Then, for observationally equivalent sensors/outputs $i,j$, $(A,\overline{C}_i)$ and $(A,\overline{C}_j)$ are observable, but $(A,\overline{C}_{i,j})$ is not observable. Consider the system digraph $\mc{G}_A=\{\mc{V},\mc{E}\}$ with  set $\mc{V}=\{1,\dots,n\}$ denoting the state nodes and link set $\mc{E}$ denoting the state interactions. Define a \textit{contraction}, $\mc{C}_l$, as the set of nodes such that $|\mc{N}(\mc{C}_l)|<|\mc{C}_l|$, where 
$\mc{N}(\mc{C}_l)=\{j|(i,j)\in \mc{E}, i \in \mc{C}_l\}$ \cite{icassp2016,murota}. Define an SCC, $\mc{S}_l$, as the set of nodes $i,j\in\mc{V}$ such that $i\overset{path}{\longrightarrow}j$ and $j\overset{path}{\longrightarrow}i$. A \textit{parent SCC} is, then, defined as an SCC with no outgoing link to any other SCC \cite{icassp2016}. Output of parent SCCs are known to recover the  \textit{output-connectivity} of $\mc{G}_A$, and output of contractions are known to recover the \textit{cyclicity} of $\mc{G}_A$ and  rank of the system matrix $A$ \cite{icassp2016}.
\begin{lem}[\cite{jstsp,acc13_kar}] \label{lem_obs2}
    Given system digraph $\mc{G}_A$, outputs of one state node in every parent SCC $\mc{S}_l$ and every contraction $\mc{C}_l$ are sufficient for structural $(A,C)$-observability, i.e., to satisfy Assumption~\ref{ass_obs}. 
\end{lem} 

The above lemma implies that the set of state nodes in a contraction $\mc{C}_l$ and a parent SCC $\mc{S}_l$ in $\mc{G}_A$ are observationally equivalent.
In this direction, $Q+1$ different state-outputs from every contraction and parent-SCC are sufficient for $Q$-redundant observability. We assign these outputs to $Q+1$ sensors. Then, removing any $Q$ sensors, the remaining ones include (at least) one output from every $\mc{S}_l$ and $\mc{C}_l$. Thus, from Lemma~\ref{lem_obs2},
the pair $(A,\overline{C}_Q)$ (with $\overline{C}_Q$ as the output matrix after removal of any $Q$ rows) remains observable.  

\subsection{$Q$-redundant distributed observer} \label{sec_Q2}
Given the outputs and sensors for $Q$-redundant observability, this section provides the $Q$-redundant distributed observer. 
Following Lemma~\ref{lem_dist_obs},  we improve the network-connectivity of $\mc{G}_\beta$ and $\mc{G}_\alpha$ to gain $Q$-redundant $(W\otimes A, D_C)$-observability as follows, 
\begin{enumerate}
    \item Design $\mc{G}_\beta$ to be $Q$-vertex-connected, e.g., via computationally-efficient algorithms in \cite{wu2008construction}. Recall that a $Q$-vertex-connected graph \textit{remains strongly-connected after cutting any $Q$ (sensor) nodes from the network.} 
    \item For the \textit{hub-network} $\mc{G}_\alpha$, following  Lemma~\ref{lem_dist_obs}, every $\alpha$-sensor $i$ (say $\alpha_1$) as a network hub, directly shares its output with every other sensor $j$  not observationally equivalent with $i$, (all other sensors except $\alpha_2$).   
\end{enumerate}
The above connectivity ensures that every sensor directly links from $Q+1$ set of  $\alpha$-sensors and $Q+1$ paths to and from other sensors.
Then,  protocol \eqref{eq_p}-\eqref{eq_m} over these $Q$-redundant networks $\mc{G}_\beta$ and  $\mc{G}_\alpha$ can tolerate failure/isolation of any $Q$ sensors without losing distributed observability over the network (from Lemma~\ref{lem_dist_obs}), i.e., a $Q$-redundant distributed estimator. In similar setup, $Q$-edge-connectivity can be used for survivable network design \cite{jabal2021approximation}
resilient to link removal.   

\subsection{Illustrative Example} \label{sec_ex}
\begin{figure}
	\centering
	\includegraphics[width=1.8in]{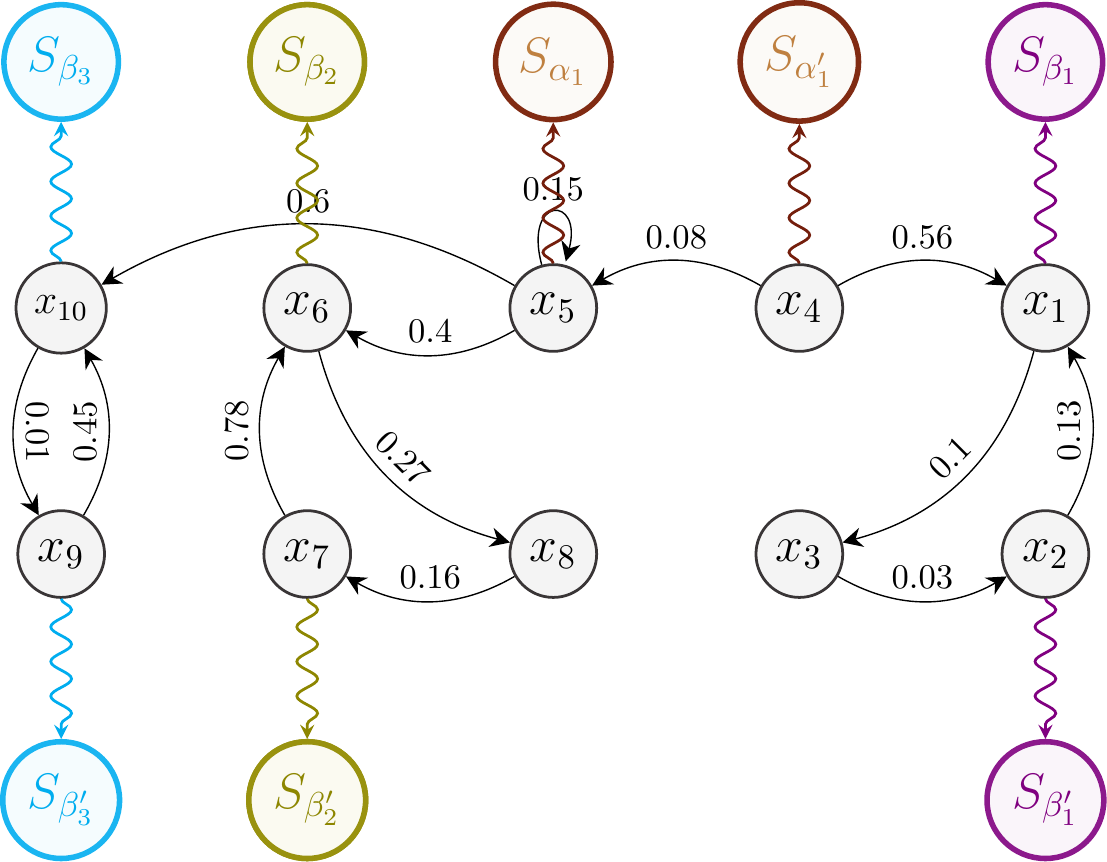}
	\includegraphics[width=1.45in]{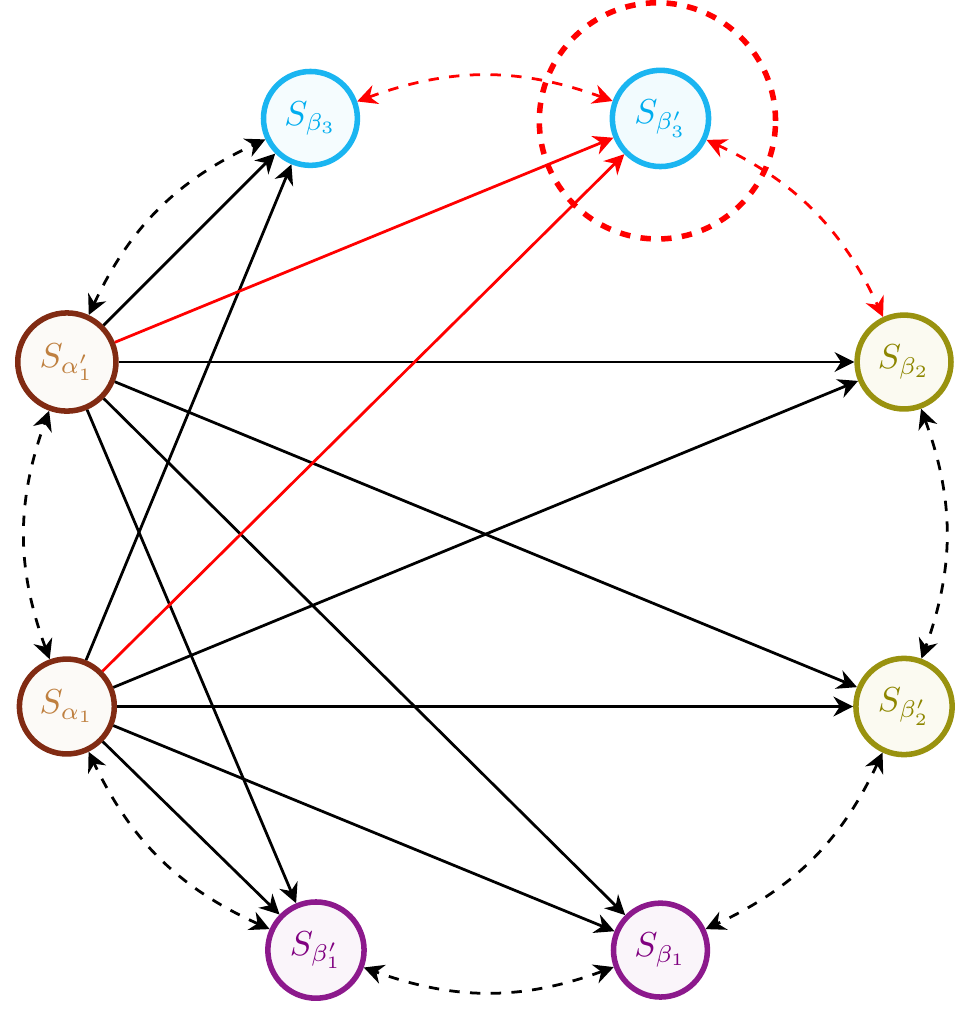}
	\caption{ (Left) This figure shows an example system digraph $\mc{G}_A$ (gray nodes) and    sensors with outputs of state nodes in $4$ parent SCCs and contraction. A link from, e.g., node $x_1$ to $x_3$ with weight $0.1$ in $\mc{G}_A$ implies that $A_{31}=0.1$. Sensors of the same color (e.g., blue colored $\beta_3$ and $\beta'_3$) are observationally equivalent with outputs of the same component (SCC or contraction). This implies that by isolating/removing one faulty output, the system remains observable via the other ones. For example, removing $\beta'_3$, $\mc{G}_A$ is globally observable to the remaining $7$ sensors. 
	(Right) This figure shows an example sensor network with solid links as network $\mc{G}_\alpha$ ($\alpha$ sensors as the hubs) and dashed-links as an example $1$-connected SC network $\mc{G}_\beta$. Both $\mc{G}_\beta$ and $\mc{G}_\alpha$ include self-loops (not shown for simplicity). By this setup, $(W \otimes A,D_C)$ is $1$-redundant observable, implying that by isolating any one sensor, e.g., $\beta'_3$
	(shown via red dashed circle) and removing its incoming/outgoing links (red-colored), $(W \otimes A,D_C)$ remains observable and  sufficient conditions in Lemmas~\ref{lem_dist_obs} and~\ref{lem_obs2} hold.} 
	\label{fig_Q} \vspace{-0.55cm}
\end{figure}

To illustrate the $Q$-redundant results, consider the system digraph $\mc{G}_A$ in Fig.~\ref{fig_Q}(Left), containing three parent SCCs $\mc{S}_1=\{x_1,x_2,x_3\}, \mc{S}_2=\{x_6,x_7,x_8\}, \mc{S}_3=\{x_9,x_{10}\}$ and one contraction $\mc{C}_1=\{x_4,x_5,x_2,x_7,x_9\}$. 
We aim to design a $1$-redundant distributed observer, i.e., an observer robust to removal/isolation of any one output (sensor). Following Section~\ref{sec_Q1}, taking $4$ outputs from $\{x_1,x_5,x_6,x_9\}$ the system digraph is structurally observable. By adding another  set of $4$ outputs from states $\{x_2,x_4,x_7,x_{10}\}$, the system is (structurally) $1$-redundant observable. Following Section~\ref{sec_Q2}, we design $\mc{G}_\beta$ as the $2$-vertex-connected graph and $\mc{G}_\alpha$ as a hub-network shown in Fig.~\ref{fig_Q}(Right) to gain $1$-redundant distributed observability. Then, in protocol \eqref{eq_p}-\eqref{eq_m}, matrix $W$ (as the adjacency of $\mc{G}_\beta$) is designed row-stochastic; for example, by considering random positive non-zero entries, and then, dividing each row by the row-sum. Matrix $U$ is the 0-1 adjacency matrix of $\mc{G}_\alpha$. Then, block-diagonal $K$ can be designed via Algorithm~\ref{alg_lmi}. This distributed observer is robust to failure of $1$ sensor, or, more precisely, to failure of $1$ output from every parent SCC and contraction.     

\section{Simulation} \label{sec_sim}
\begin{figure}
	\centering
	\includegraphics[width=1.7in]{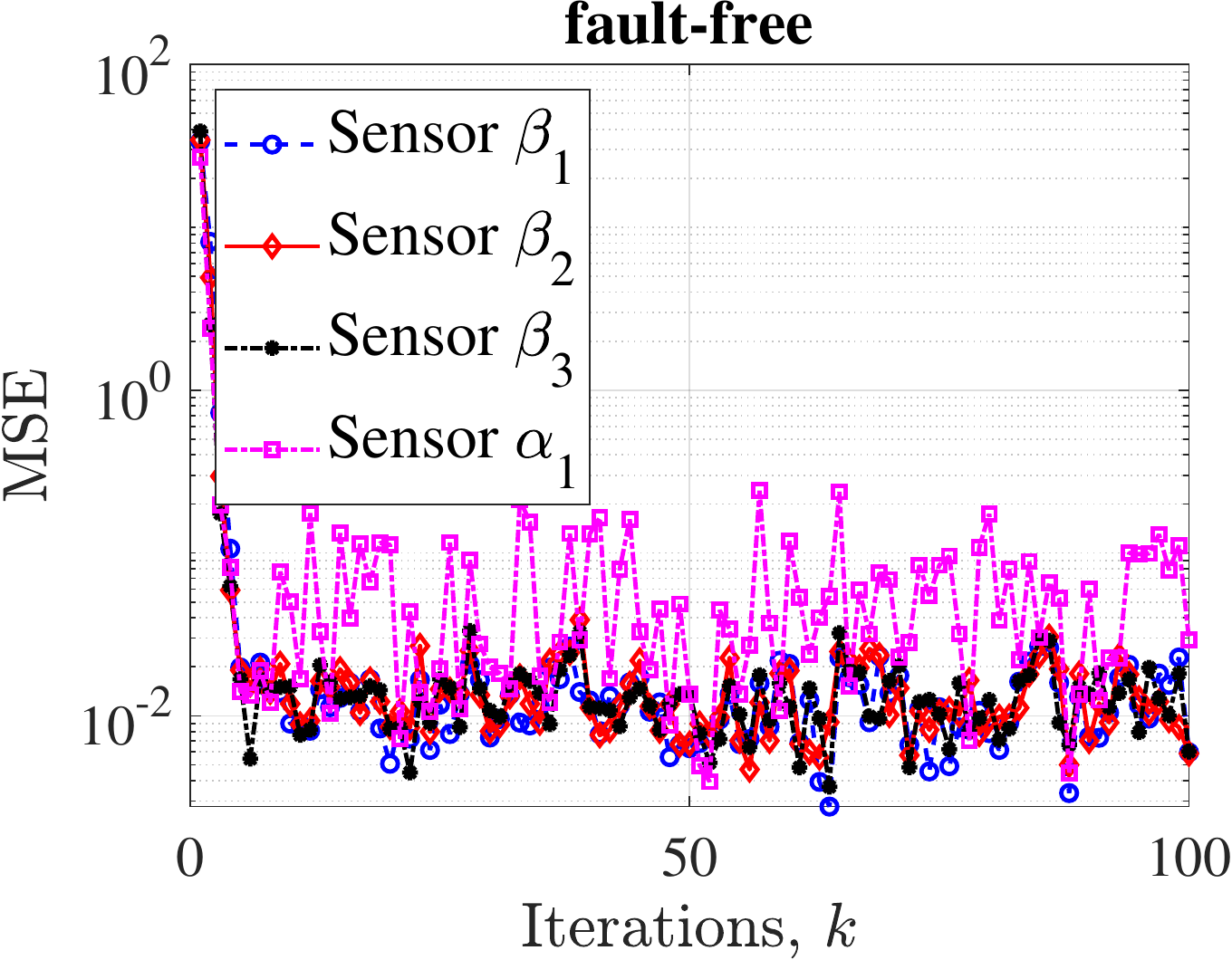}
	\includegraphics[width=1.64in]{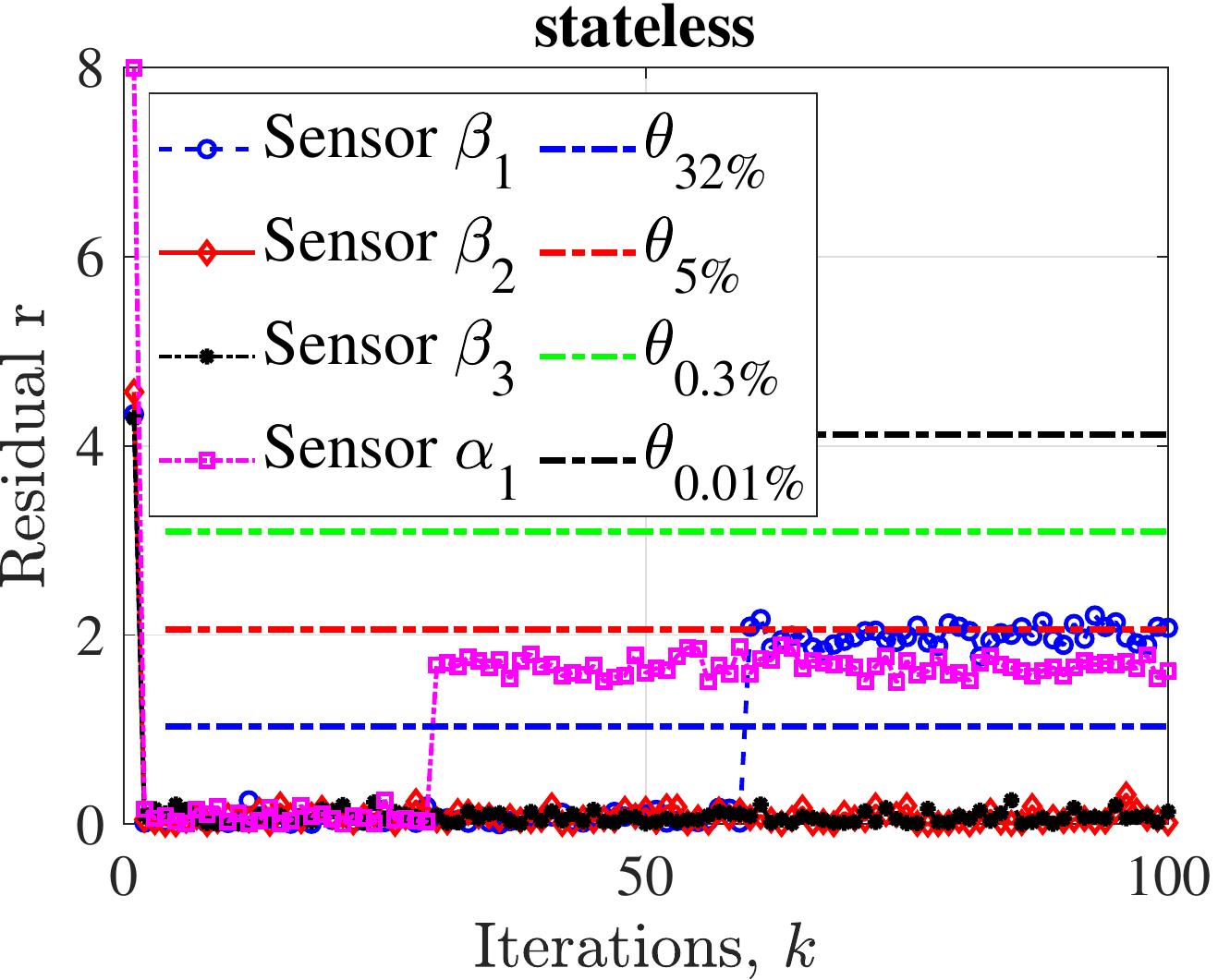}
	\includegraphics[width=1.7in]{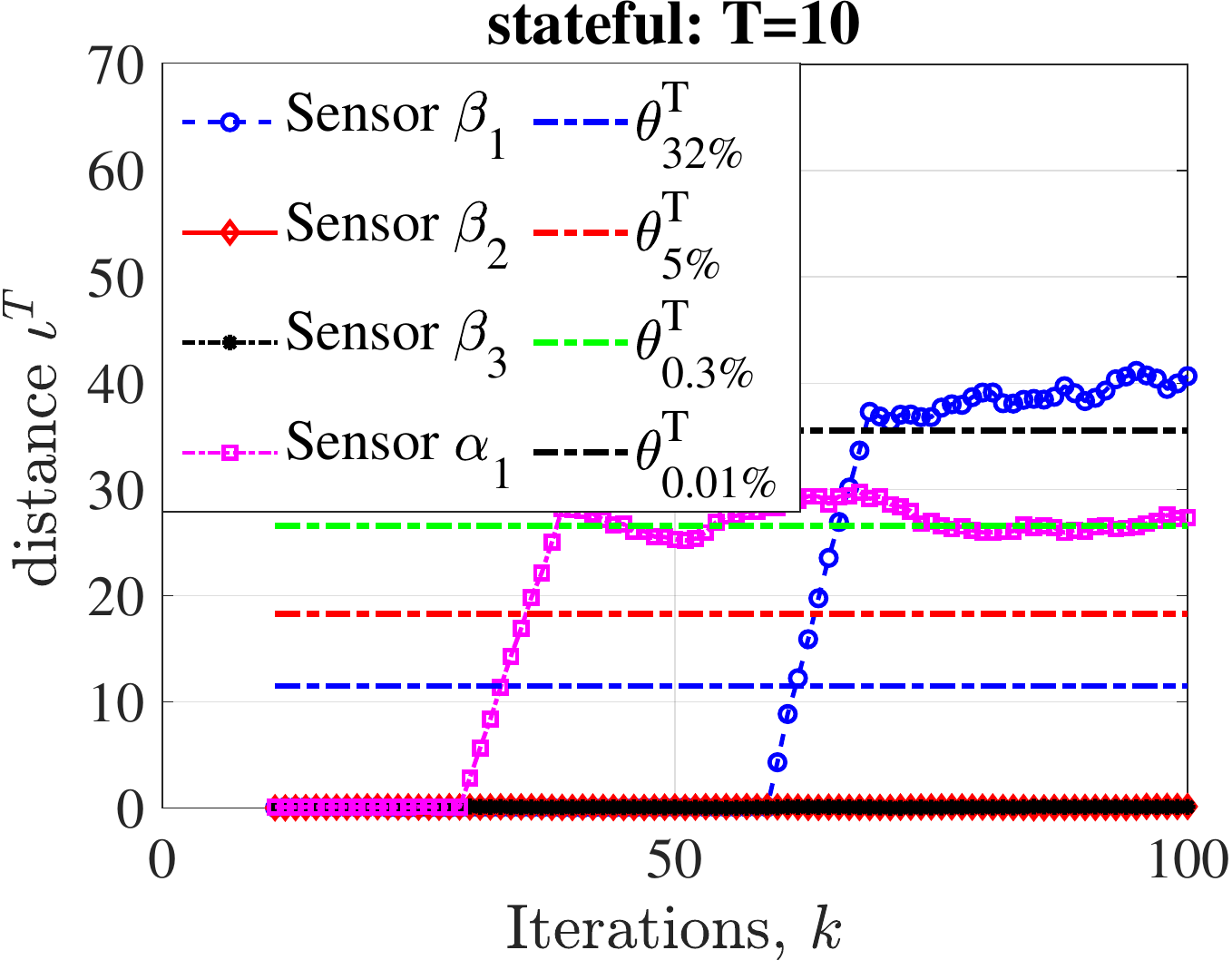}
	\includegraphics[width=1.64in]{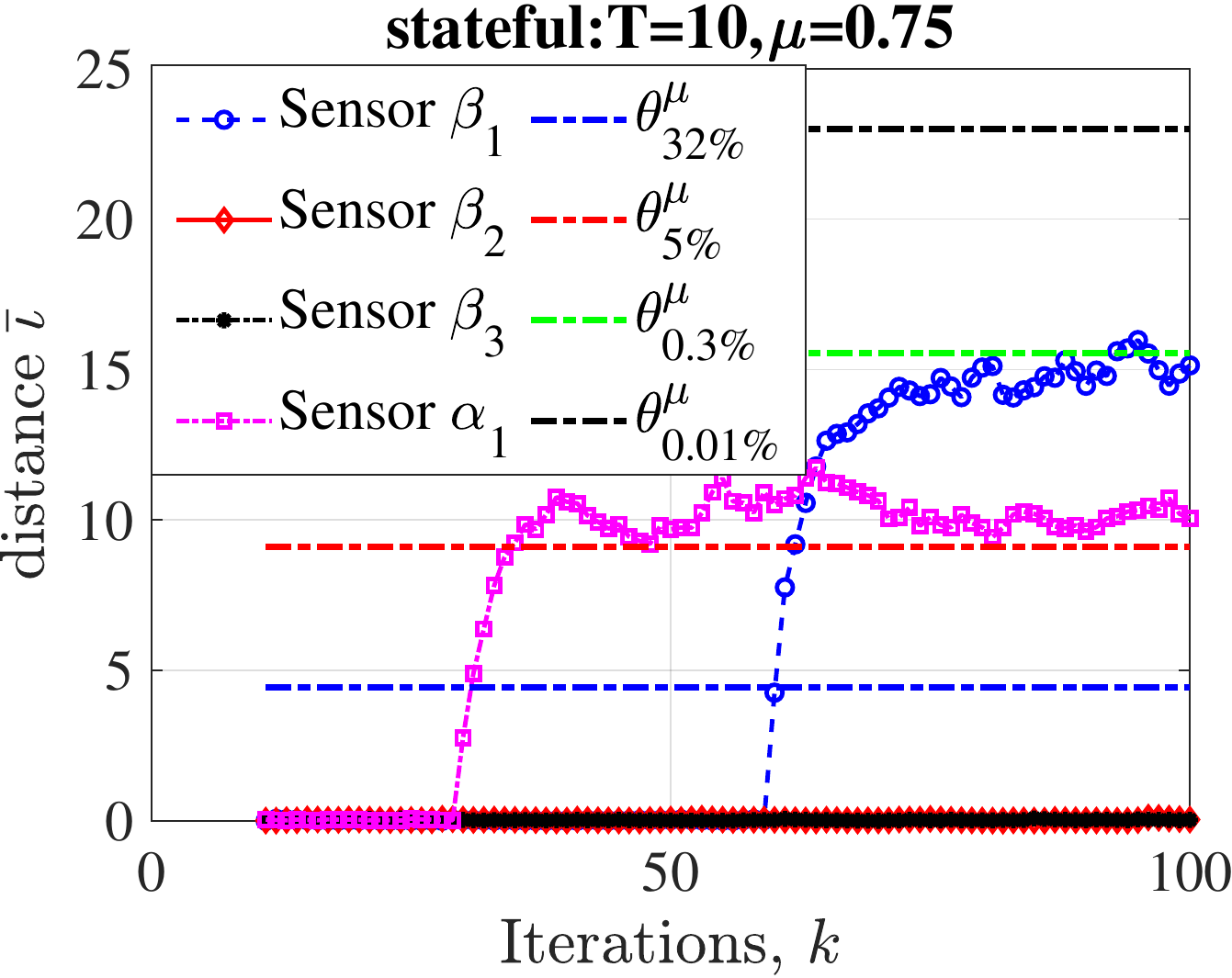}	
	\caption{This figure shows (TopLeft) the bounded-MSE performance of the proposed protocol \eqref{eq_p}-\eqref{eq_m} in the \textit{absence} of output bias, i.e., $f_i=0,\forall i$, and \textit{localized detection} in the \textit{presence} of output bias, $f_{\beta_1}(k\geq 60)=2$ and $f_{\alpha_1}(k\geq 30) \sim \mc{N}(2,0.5)$): (TopRight) stateless case considering no history of the residuals, (BelowLeft) stateful case over time-window $T=10$ and equally-weighting all residual history,  (BelowRight) stateful case over the same time-window and weighting the residual history by $\mu=0.75$.         } 
	\label{fig_sim1} \vspace{-0.45cm}
\end{figure}

For simulation, we consider the same example in Section~\ref{sec_ex} representing a rank deficient system of $n=10$ nodes with non-zero entries as the given link weights in Fig.~\ref{fig_Q}(Left). We consider $N=4$ outputs from $\{x_1,x_6,x_9,x_5\}$ associated with sensors $\{\beta_1,\beta_2,\beta_3,\alpha_1\}$, satisfying Lemma~\ref{lem_obs2} and Assumption~\ref{ass_obs}. Network $\mc{G}_\alpha$ includes direct links (with weight $1$) from $\alpha_1$ to $\beta_1,\beta_2,\beta_3$, and $\mc{G}_\beta$ is considered as a directed cycle $1\rightarrow 4 \rightarrow 3 \rightarrow 2 \rightarrow 1$ with random row-stochastic link weights. 
The non-zero entries of the output matrix $C$ are set equal to $1$. The gain matrix $K$ is designed via Algorithm~\ref{alg_lmi} with $\epsilon=0.14$. The mean-square error (MSE) under the proposed protocol \eqref{eq_p}-\eqref{eq_m} is bounded steady-state as shown in Fig.~\ref{fig_sim1}(TopLeft) with $\nu\sim \mc{N}(0,0.01)$ and $\zeta \sim \mc{N}(0,0.01)$. To check the performance in the presence of faults or anomalies, we consider two bias $f_{\beta_1}(k\geq 60)=2$ and $f_{\alpha_1}(k\geq 30) \sim \mc{N}(2,0.5)$. Our localized (or distributed) detection logic follows algorithm~\ref{alg_logic} for FAR $p=32\%, 5\%, 0.3\%, 0.01\%$. For the stateless case, the residual $r_i$ at the faulty sensors $\beta_1$ and $\alpha_1$ are between the thresholds $\theta_{32\%}$ and $\theta_{5\%}$. Therefore, the local detector declares the fault probability of more than $68\%$ and less than $95\%$ at these sensors and no fault at the other two. For the stateful case, considering the distance measure $\iota_i^T$ in \eqref{eq_z} with $T=10$-steps  and thresholds \eqref{eq_theta_T}, the local detector declares higher probabilities of $99.7\%$ and $99.99\%$ respectively at sensors $\alpha_1$ and $\beta_1$, however \textit{with certain delay} (about $10$-steps delay as in Fig.~\ref{fig_sim1}(BelowLeft)). One can reduce this alarm delay  by weighting the residual history by $\mu$ as in \eqref{eq_z2}, however via thresholds of higher FAR. For $\mu=0.75$, the weighted distance measure $\overline{\iota}_i$ and the thresholds \eqref{eq_theta_T2} are shown in Fig.~\ref{fig_sim1}(BelowRight). The detection probability with this logic is approximately $99.7\%$ at sensor $\alpha_1$ and  $95\%$ at sensor $\beta_1$, with alarm delay reduced to $7$ time-steps.       

\begin{figure} [h]
	\centering
	\includegraphics[width=1.58in]{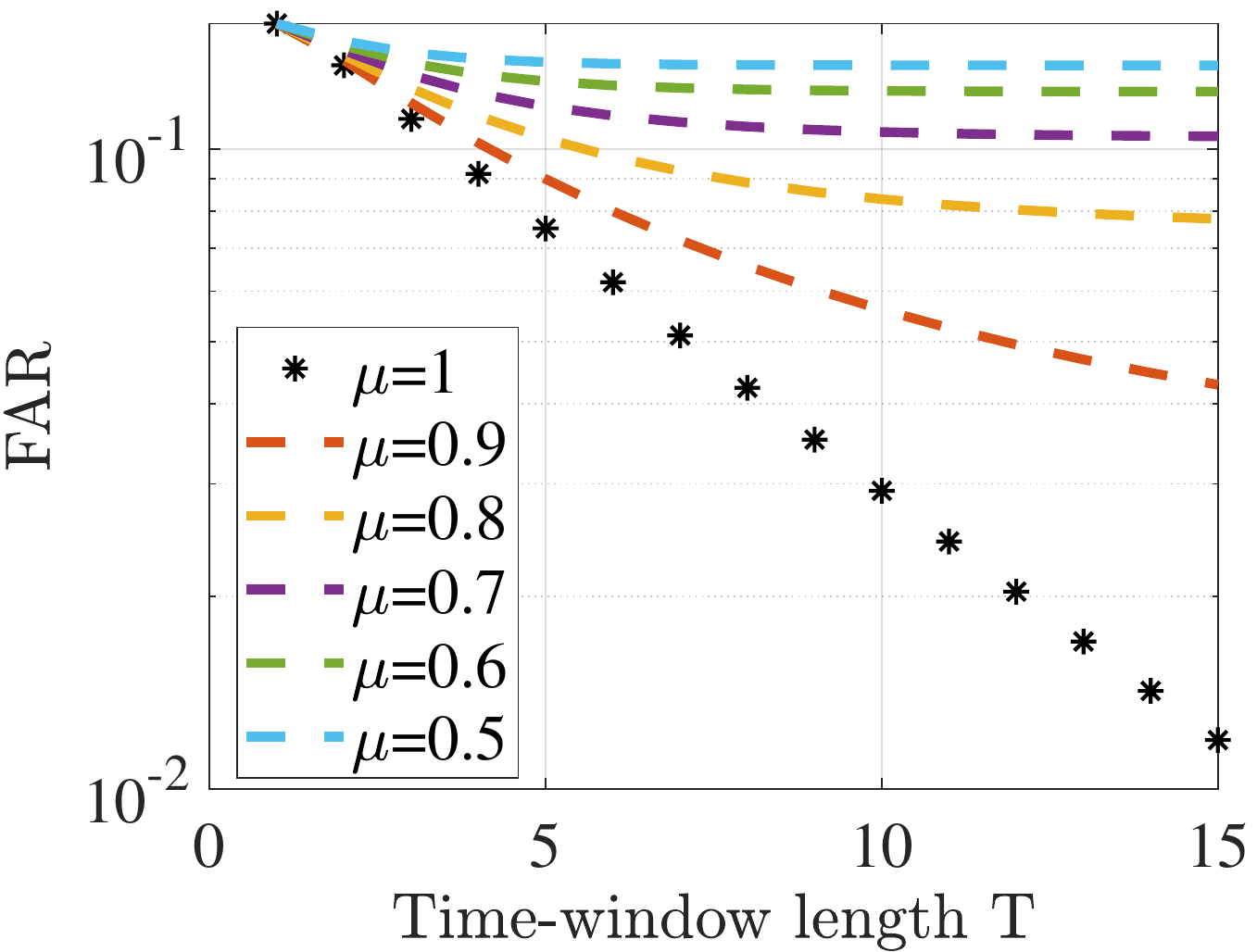}
	\includegraphics[width=1.55in]{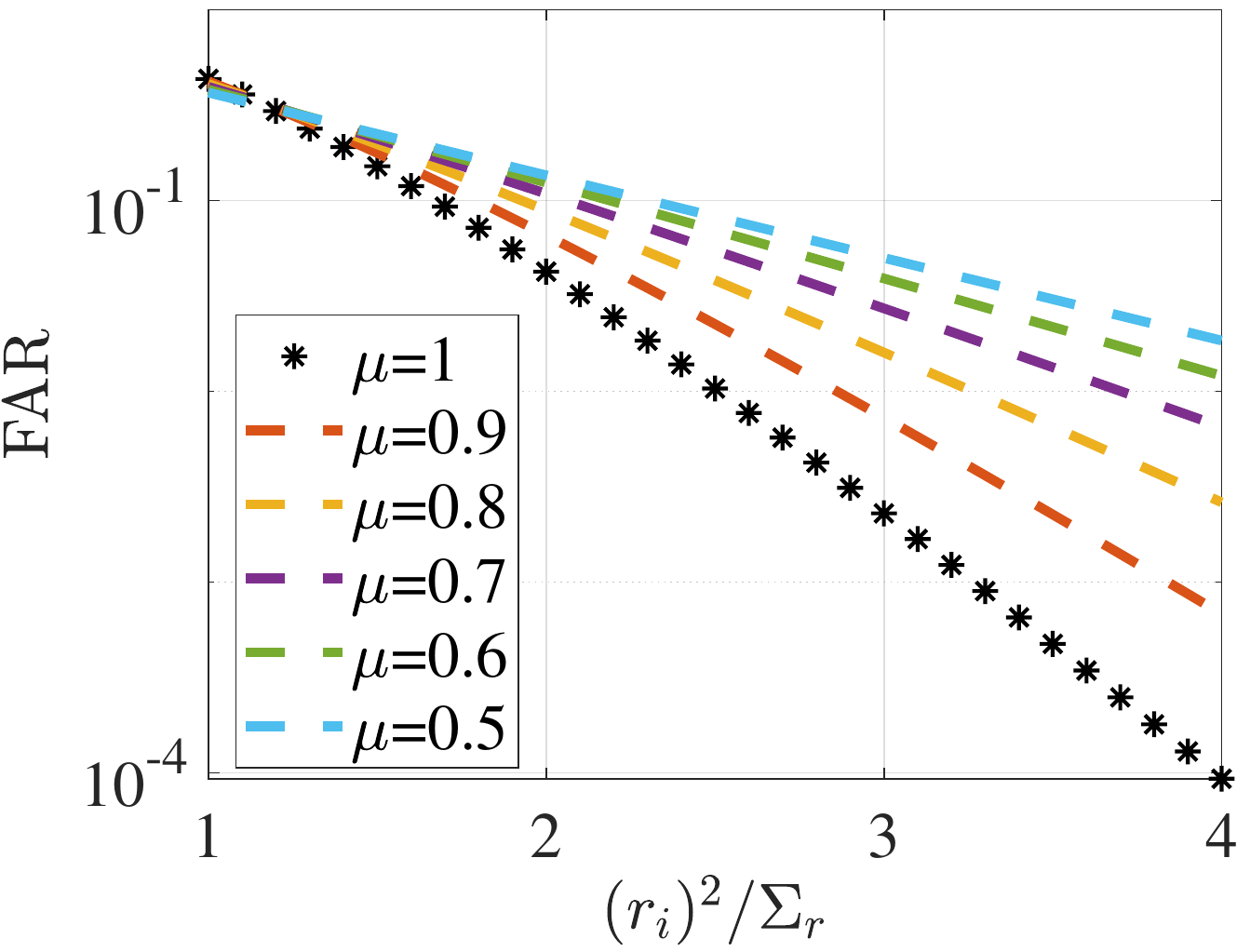}
	\caption{(Left) FAR increases for \textit{some} shorter time-windows $T$ and smaller $\mu$ (for  normalized residual $\frac{r_i^2}{\Sigma_r^i}=2$). (Right) FAR increases for larger values of normalized residual 
	 and smaller values of $\mu$ in \eqref{eq_z2} (for fixed $T=8$). } \label{fig_Tmu}
	 \vspace{-0.4cm}
\end{figure}

Next, we compare the performance of the stateful detectors for different values of $T$ and $\mu$. Consider a constant fault resulting in biased  residual $r_i^2=2\Sigma_r^i$. For the stateless detector, the associated FAR is equal to $5\%$. For the stateful case, the threshold's FAR can be defined via \eqref{eq_p_Tmu}, shown in Fig.~\ref{fig_Tmu}(Left) for different  $T$ and $\mu$ values. Clearly, for greater $\mu$,  the FAR is lower; however, as discussed in Remark~\ref{rem_mu} and Fig.~\ref{fig_sim1}, large $\mu$ values result in longer delays to raise the alarm. Further, for smaller values, e.g., $\mu=0.5,0.6$,  the FAR is almost constant for $T\geq8$ and, thus, longer time windows do not improve the FAR.   
Fig.~\ref{fig_Tmu}(Right) shows the FAR  versus  $\frac{r_i^2}{\Sigma_r^i}$ for different $\mu$ values (for $T=8$). It is clear that as $\mu \rightarrow 1$ the FAR decreases (with $\mu=1$ giving \eqref{eq_theta_T}).           

\section{Conclusion and Future Works} \label{sec_con}
Stateful and stateless local detection mechanisms over distributed estimation networks are considered with a trade-off between alarm delay and FAR. The stateful case shows lower FAR with possibly delayed alarm over long sliding time windows. The solutions are of polynomial-order complexity; for example, system digraph decomposition into parent SCCs and contractions 
is of complexity $\mc{O}(n^{2.5})$ \cite{murota}.  
As future research, one may consider  possible time-delay in the communication network \cite{delay_arxiv2021,hadjicostis2013average}.

\bibliographystyle{IEEEbib}
\bibliography{bibliography}

\begin{thebibliography}{10}

\bibitem{csl_iot}
M.~Doostmohammadian and H.~R. Rabiee,
\newblock ``On the observability and controllability of large-scale {IoT}
  networks: Reducing number of unmatched nodes via link addition,''
\newblock {\em IEEE Control Systems Letters}, vol. 5, no. 5, pp. 1747--1752,
  2021.

\bibitem{chen2018internet}
Y.~Chen, S.~Kar, and J.~M.~F. Moura,
\newblock ``The internet of things: Secure distributed inference,''
\newblock {\em IEEE Signal Proc. Mag.}, vol. 35, no. 5, pp. 64--75, 2018.

\bibitem{icas21_social}
M.~Doostmohammadian, T.~Charalambous, M.~Shafie-khah, N.~Meskin, and U.~A.
  Khan,
\newblock ``Simultaneous distributed estimation and attack detection/isolation
  in social networks: Structural observability, kronecker-product network, and
  chi-square detector,''
\newblock in {\em IEEE International Conference on Autonomous Systems}, 2021,
  pp. 344--348.

\bibitem{camsap11}
U.~A. Khan and M.~Doostmohammadian,
\newblock ``A sensor placement and network design paradigm for future smart
  grids,''
\newblock in {\em IEEE Workshop on Computational Adv. in Multi-Sensor Adap.
  Proc.}, 2011, pp. 137--140.

\bibitem{ilic2010modeling}
M.~D. Ili{\'c}, L.~Xie, U.~A. Khan, and J.~M.~F. Moura,
\newblock ``Modeling of future cyber--physical energy systems for distributed
  sensing and control,''
\newblock {\em IEEE Transactions on Systems, Man, and Cybernetics-Part A:
  Systems and Humans}, vol. 40, no. 4, pp. 825--838, 2010.

\bibitem{umsonst2019tuning}
D.~Umsonst,
\newblock {\em Tuning of anomaly detectors in the presence of sensor attacks},
\newblock Ph.D. thesis, KTH Royal Institute of Technology, 2019.

\bibitem{abbaszadeh2019system}
M.~Abbaszadeh,
\newblock ``System and method for anomaly and cyber-threat detection in a wind
  turbine,'' 2019,
\newblock US Patent App. 15/988,515.

\bibitem{davoodi2016event}
M.~Davoodi, N.~Meskin, and K.~Khorasani,
\newblock ``Event-triggered multiobjective control and fault diagnosis: A
  unified framework,''
\newblock {\em IEEE Trans. on Industrial Informatics}, vol. 13, no. 1, pp.
  298--311, 2016.

\bibitem{navi2018sensor}
M.~Navi, N.~Meskin, and M.~Davoodi,
\newblock ``Sensor fault detection and isolation of an industrial gas turbine
  using partial adaptive {KPCA},''
\newblock {\em Journal of Process Control}, vol. 64, pp. 37--48, 2018.

\bibitem{he2020secure}
X.~He, X.~Ren, H.~Sandberg, and K.~H Johansson,
\newblock ``How to secure distributed filters under sensor attacks?,''
\newblock {\em IEEE Transactions on Automatic Control, arXiv preprint
  arXiv:2004.05409}, 2021.

\bibitem{battilotti2021stability}
S.~Battilotti, F.~Cacace, and M.~d’Angelo,
\newblock ``A stability with optimality analysis of consensus-based distributed
  filters for discrete-time linear systems,''
\newblock {\em Automatica}, vol. 129, pp. 109589, 2021.

\bibitem{tcns20}
M.~Doostmohammadian and N.~Meskin,
\newblock ``Sensor fault detection and isolation via networked estimation:
  Full-rank dynamical systems,''
\newblock {\em IEEE Trans. Control of Net. Systems}, vol. 8, no. 2, pp.
  987--996, 2021.

\bibitem{morse2019cdc}
L.~Wang, J.~Liu, A.~S. Morse, and B.~D.~O. Anderson,
\newblock ``A distributed observer for a discrete-time linear system,''
\newblock in {\em 58th IEEE Conference on Decision and Control}, 2019, pp.
  367--372.

\bibitem{zamani2018distributed}
D.~Marelli, M.~Zamani, M.~Fu, and B.~Ninness,
\newblock ``Distributed kalman filter in a network of linear systems,''
\newblock {\em Systems \& Control Letters}, vol. 116, pp. 71--77, 2018.

\bibitem{jstsp}
M.~Doostmohammadian and U.~Khan,
\newblock ``On the genericity properties in distributed estimation: Topology
  design and sensor placement,''
\newblock {\em IEEE J. of Sel. Topics in Signal Processing}, vol. 7, no. 2, pp.
  195--204, 2013.

\bibitem{giraldo2018survey}
J.~Giraldo, D.~Urbina, A.~Cardenas, J.~Valente, M.~Faisal, J.~Ruths, N.~O.
  Tippenhauer, H.~Sandberg, and R.~Candell,
\newblock ``A survey of physics-based attack detection in cyber-physical
  systems,''
\newblock {\em ACM Computing Surveys (CSUR)}, vol. 51, no. 4, pp. 1--36, 2018.

\bibitem{rikos2020privacy}
A.~I. Rikos, T.~Charalambous, K.~H. Johansson, and C.~N. Hadjicostis,
\newblock ``Privacy-preserving event-triggered quantized average consensus,''
\newblock in {\em IEEE Conference on Decision and Control}, 2020, pp.
  6246--6253.

\bibitem{sundaram_2021resilient}
A.~Mitra, F.~Ghawash, S.~Sundaram, and W.~Abbas,
\newblock ``On the impacts of redundancy, diversity, and trust in resilient
  distributed state estimation,''
\newblock {\em IEEE Transactions on Control of Network Systems}, 2021.

\bibitem{kim2018detection}
J.~Kim, C.~Lee, H.~Shim, Y.~Eun, and J.~H. Seo,
\newblock ``Detection of sensor attack and resilient state estimation for
  uniformly observable nonlinear systems having redundant sensors,''
\newblock {\em IEEE Transactions on Automatic Control}, vol. 64, no. 3, pp.
  1162--1169, 2018.

\bibitem{dsvm}
M.~Doostmohammadian, A.~Aghasi, T.~Charalambous, and U.~Khan,
\newblock ``Distributed support vector machines over dynamic balanced directed
  networks,''
\newblock {\em IEEE Control Systems Let.}, vol. 6, pp. 758 -- 763, 2021.

\bibitem{asilomar14}
M.~Doostmohammadian and U.~A. Khan,
\newblock ``Vulnerability of {CPS} inference to {DoS} attacks,''
\newblock in {\em 48th Annual Asilomar Conference on Signals, Systems, and
  Computers}, 2014, pp. 2015--2018.

\bibitem{GRACY2021109925}
S.~Gracy, J.~Milošević, and H.~Sandberg,
\newblock ``Security index based on perfectly undetectable attacks:
  Graph-theoretic conditions,''
\newblock {\em Automatica}, vol. 134, pp. 109925, 2021.

\bibitem{guichard2017introduction}
D.~Guichard,
\newblock ``An introduction to combinatorics and graph theory,''
\newblock {\em Whitman College-Creative Commons}, 2017.

\bibitem{acc13_kar}
S.~Pequito, S.~Kar, and A.~P. Aguiar,
\newblock ``A structured systems approach for optimal actuator-sensor placement
  in linear time-invariant systems,''
\newblock in {\em American Control Conference}, 2013, pp. 6123--6128.

\bibitem{icassp2016}
M.~Doostmohammadian and U.~A. Khan,
\newblock ``Measurement partitioning and observational equivalence in state
  estimation,''
\newblock in {\em IEEE Conference on Acoustics, Speech and Signal Processing},
  2016, pp. 4855--4859.

\bibitem{usman_cdc:11}
U.~A. Khan and A.~Jadbabaie,
\newblock ``Coordinated networked estimation strategies using structured
  systems theory,''
\newblock in {\em 49th IEEE Conference on Decision and Control}, Orlando, FL,
  Dec. 2011, pp. 2112--2117.

\bibitem{khan2014collaborative}
U.~A. Khan and A.~Jadbabaie,
\newblock ``Collaborative scalar-gain estimators for potentially unstable
  social dynamics with limited communication,''
\newblock {\em Automatica}, vol. 50, no. 7, pp. 1909--1914, 2014.

\bibitem{Bausch2013OnTE}
J.~Bausch,
\newblock ``On the efficient calculation of a linear combination of chi-square
  random variables with an application in counting string vacua,''
\newblock {\em Journal of Physics A}, vol. 46, pp. 505202, 2013.

\bibitem{icas21_contraction}
M.~Doostmohammadian, T.~Charalambous, M.~Shafie-khah, H.~R. Rabiee, and U.~A.
  Khan,
\newblock ``Analysis of contractions in system graphs: Application to state
  estimation,''
\newblock in {\em IEEE International Conference on Autonomous Systems}, 2021,
  pp. 359--363.

\bibitem{murota}
K.~Murota,
\newblock {\em Matrices \& matroids for systems analysis},
\newblock Springer, 2000.

\bibitem{wu2008construction}
Y.~Wu and Y.~Li,
\newblock ``Construction algorithms for k-connected m-dominating sets in
  wireless sensor networks,''
\newblock in {\em 9th ACM symposium on Mobile ad hoc networking and computing},
  2008, pp. 83--90.

\bibitem{jabal2021approximation}
A.~Jabal~Ameli,
\newblock {\em Approximation algorithms for survivable network design},
\newblock Ph.D. thesis, Universit{\`a} della Svizzera italiana, 2021.

\bibitem{delay_arxiv2021}
M.~Doostmohammadian, M.~Pirani, U.~A. Khan, and T.~Charalambous,
\newblock ``Consensus-based distributed estimation in the presence of
  heterogeneous, time-invariant delays,''
\newblock {\em IEEE Control Systems Letters}, vol. 6, pp. 1598 -- 1603, 2021.

\bibitem{hadjicostis2013average}
C.~N. Hadjicostis and T.~Charalambous,
\newblock ``Average consensus in the presence of delays in directed graph
  topologies,''
\newblock {\em IEEE Transactions on Automatic Control}, vol. 59, no. 3, pp.
  763--768, 2013.

\end{thebibliography}
\end{document}